\documentclass[11pt]{article}
\usepackage{graphicx}
\usepackage{indentfirst,csquotes}

\usepackage{amssymb,amsthm,amsmath}
\usepackage{xcolor,paralist,hyperref,titlesec,fancyhdr,etoolbox}

\usepackage{tikz}
\def\checkmark{\tikz\fill[scale=0.4](0,.35) -- (.25,0) -- (1,.7) -- (.25,.15) -- cycle;} 
\usepackage{multirow}
\usepackage{xcolor}
\usepackage{booktabs}
\usepackage{hyperref}
\usepackage{natbib}
\usepackage{graphicx}
\usepackage{adjustbox}
\usepackage{hyperref}
\usepackage{seqsplit}

\usepackage{subfig}
\usepackage{graphicx}
\usepackage{csquotes}
\usepackage{authblk} 

\usepackage{url}
\begin{document}

\title{Banal Deception \& Human-AI Ecosystems: A Study of People's Perceptions of LLM-generated Deceptive Behaviour}
\author[1]{Xiao Zhan\thanks{Both authors contributed equally to this research.}}
\author[2]{Yifan Xu\protect\footnotemark[1]} 
\author[3]{Noura Abdi}
\author[4]{Joe Collenette}
\author[1]{Ruba Abu-Salma}
\author[1]{Ștefan Sarkadi}

\affil[1]{King's College London, \{xiao.zhan, ruba.abu-salma, stefan.sarkadi\}@kcl.ac.uk}
\affil[2]{The University of Manchester, yifan.xu@manchester.ac.uk}
\affil[3]{Liverpool Hope University, abdin@hope.ac.uk}
\affil[4]{The University of Chester, j.collenette@chester.ac.uk}


\date{} 

\maketitle

\begin{abstract}
Large language models (LLMs) can provide users with false, inaccurate, or misleading information, and we consider the output of this type information as what \cite{natale2021deceitful} calls `banal' deceptive behaviour. Here, we investigate peoples' perceptions of ChatGPT-generated deceptive behaviour and how this affects peoples' own behaviour and trust. To do this, we use a mixed-methods approach comprising of (i) an online survey with 220 participants and (ii) semi-structured interviews with 12 participants. Our results show that (i) the most common types of deceptive information encountered were over-simplifications and outdated information; (ii) humans' perceptions of trust and `worthiness' of talking to ChatGPT are impacted by `banal' deceptive behaviour; (iii) the perceived responsibility for deception is influenced by education level and the frequency of deceptive information; and (iv) users become more cautious after encountering deceptive information, but they come to trust the technology more when they identify advantages of using it. Our findings contribute to the understanding of human-AI interaction dynamics in the context of \textit{Deceptive AI Ecosystems}, and highlight the importance of user-centric approaches to mitigating the potential harms of deceptive AI technologies.

\end{abstract}
\section{Introduction}
\label{Introduction}


\begin{quote}
    
    \textit{``A deceptive AI ecosystem represents not just the technical aspects of developing deceptive AI technologies, but how the societal and evolutionary pressures influence human interaction with these technologies as individuals, groups, and organizations. This creates an ever-evolving informational feedback loop between hybrid societies where humans and machines communicate as agents and the emerging socio-economical regulatory norms, human and societal values, business decisions, power structures, communication about AI technologies, and market behaviour."}\citep{sarkadi2023deceptive}
\end{quote}

In the past year, ChatGPT\footnote{\url{https://openai.com/chatgpt/}} has emerged as a powerful conversational AI system that has captured the attention of researchers and users alike. This advanced generative AI system is designed to generate human-like responses to user queries, making it an attractive tool for a range of applications, including customer service~\citep{chatgpt-customer}, and education~\citep{kohnke2023chatgpt,xiao2023exploratory}. Despite ChatGPT's remarkable performance, its potential for generating deceptive information in its responses to user prompts should not be ignored. However, erroneous information provided by ChatGPT can lead to undesirable outcomes, causing users to lose trust in the system, and impede its adoption in contexts in which it might actually prove to be useful \citep{zhan2023deceptive}.

Analogous to the distinction between Strong and Weak AI, there are two types of deceptive AI technologies that can act as agents within Deceptive AI Ecosystems~\citep{sarkadi2023deceptive}. The first type comes in the form of fully autonomous AI agents whose cognitive architecture allows them to do the same thing human minds can do, which in this case is deceiving in the same way humans do. The development of the first type of deceptive AI technology follows the process that Boden described as \textit{`making computers that do the same thing minds can do...'} \citep{Boden:2016:AI}. Developing the second type of deceptive AI technologies has less to do with the cognitive capabilities of AI agents themselves and, instead, has more to do with how humans perceive AI behaviour; i.e., the effect of AI behaviour on humans' perceptions \citep{zhan2023deceptive,masters2021characterising}. The second type of deceptive AI technologies is not capable of engaging in the process of deception on its own, whether intentionally or deliberately, because such technologies lack the necessary cognitive modules and architectures enabling them to understand the meaning (semantics) or consequences of their actions in various contexts. They also lack the ability to form and use Theory of Mind \citep{verma2024theory}, which together with meta-cognition and reflection, is a necessary ingredient for deception \citep{sarkadi2021phdthesis}. Yet, their behaviour is nevertheless deceptive due to the biases of their human users and the context in which humans are interacting with them. AI researchers and engineers who build the second type of deceptive AI technologies aim to optimise the effect of AI behaviour on humans. From Natale's deceptive media perspective, this would mean that they optimise for `banal' deception \citep{natale2021deceitful}.

\textit{Banal deception} is not a process that an AI agent cognitively engages in. In this case, the AI agent does not require an ulterior motive or goal, nor the necessary cognitive capabilities to reason, plan and act to cause a desired false belief in the mind of its target. Banal deception arises from the contextual background in which the human-machine interaction takes place. Designers of technology can set up this background context in such a way that they control for banal deception by playing into humans' cognitive biases. I.e. humans become susceptible to false beliefs because inaccurate, misleading, or false informational content is provided in a context mediated by a technology that the humans interact with.

In this way, banal deception can be triggered by specific contexts that allow for the deceptive information to be presented in ways that exploit humans' cognitive biases by keeping human targets in the cognitively efficient System 1 thinking (adopting mental shortcuts in making decisions or reaching conclusions) and avoiding to trigger them into employing System 2 thinking (e.g., deliberation, argumentation, critical thinking) \citep{tversky1988rational,tversky1996reality}. This effect is also observed in human-human deception, where humans are in a truth-default state, a mental state that treats all incoming information as truthful, from which they only exit if something in the context seems `fishy' \citep{levine2019duped}.

In this paper, we are tackling the problem posed by Large Language Models (LLMs), which are successful drivers of `banal' deception that fall into the second type of deceptive AI technology. LLMs play into the tendency of humans to anthropomorphise. This tendency is driven by the cognitive biases of System 1 thinking \citep{epley2007seeing}. Moreover, due to the dynamic responses of LLM chatbot systems, and due to their adaptability to human prompts, this anthropomorphic bias can be exploited in humans with individual differences \citep{letheren2016individual}. The linguistic context of the human-LLM interaction also helps with anthropomorphisation \citep{kopp2023s}.

LLM-based chatbots, including notable examples such as OpenAI's ChatGPT, Claude\footnote{\url{https://claude.ai/}}, Microsoft's Bing Chat\footnote{\url{https://www.bing.com/}}, and Google's Bard\footnote{\url{https://bard.google.com/}}, have made significant gains on the technological market. Among these, ChatGPT has been particularly noteworthy, amassing an impressive 100 million users within just three months of its launch, thereby establishing itself as one of the fastest-growing online platforms to date~\cite{ray2023chatgpt}. These chatbots utilize the sophisticated capabilities of machine learning (ML) and natural language processing (NLP), having been extensively trained on vast datasets. This rigorous training regimen enables them to accurately decipher complex language patterns, grasp user intents, and proficiently handle intricate queries. As a result, these chatbots offer interactions that are not only more precise and refined but also capable of evolving. They leverage insights from prior interactions to continually enhance their conversational abilities, thereby setting a new benchmark in user-AI interaction efficiency. 


However, anthropomorhising these capabilities of LLMs just reinforces what is fundamentally an illusion. Not to fall for the banal deception, we need to remind ourselves that LLMs are fundamentally different from us, as  \cite{shanahan2024talking} points out. At its core, ChatGPT and LLMs are nothing other than `bullshit' machines \citep{hicks2024chatbullshit} because they lack self-awareness and knowledge about the world. Even to be able to lie you must know what you're lying about and need to be able to know the truth-value fo your statements \citep{Frankfurt-2009-OnB}. 

Shortly after its release, ChatGPT raised numerous concerns~\citep{borji2023categorical,idoracism,salilhate}, such as providing erroneous information~\citep{borji2023categorical}, exhibiting discriminatory behavior~\citep{idoracism}, and engaging in inappropriate speech and conduct~\citep{salilhate}. These concerns underscore the importance of understanding the deceptive potential of AI chatbots from the perspective of human-AI interaction. The capability of ChatGPT to provide information and interact with users, while impressive, also presents opportunities for misinformation, whether through the limitations of its training data or the inherent biases within these datasets. The nuances of deception in ChatGPT interactions, therefore, are multifaceted, stemming not only from its operational mechanics but also from how humans interpret and respond to these interactions.

This complexity of human-machine relationships leads us to our research questions, which aim to dissect the nature of deception in the interactions between users and ChatGPT and how this deception impact user behaviours and perceptions. More precisely, we aim to answer the following research questions:

\begin{description}
    \item [RQ1] What are the most common types of deceptive behavior of ChatGPT, and in which domain (e.g., research, entertainment) do they predominantly occur?
\end{description}

\begin{description}
    \item [RQ2] How do users perceive ChatGPT's worthiness and responsibility concerning deception, and what are users' behavioral responses to ChatGPT's deceptive behaviour? 
    
    \item [RQ3] How are users' perceptions (worthiness and responsibility) and behavioural responses influenced by demographics and other factors (e.g., frequency of use)?

\end{description}

We conducted \textbf{Study 1} to address RQ1 to RQ3. This study (as detailed in \S\ref{sec:study-one}) utilized an online survey administered to 220 ChatGPT users across the United Kingdom and the United States. The survey primarily focused on gathering users' responses and opinions regarding their experiences with ChatGPT, specifically targeting instances of deceptive behaviour encountered during interactions. We performed a descriptive analysis of the survey responses to answer RQ1 and RQ2. Additionally, we conducted Chi-Square tests and post-hoc analyses to examine how user perceptions are influenced by demographics and other factors, thereby addressing RQ3.

\textbf{RQ1-3 Highlights:} Our findings indicate that the most frequent types of deceptive behavior encountered by users were \textit{overly simplified} (53.64\%) and \textit{outdated information} (42.27\%), with \textit{research} being the most frequent domain for these occurrences. Our analysis shows that the \textit{frequency of deception} impacts users' perceived \textit{worthiness} of ChatGPT without being swayed by personal factors. \textit{Responsibility} for deception is influenced only by \textit{educational level} and \textit{frequency of deception}. \textit{Behavioural responses}, however, are determined by a mix of demographics (\textit{gender, age}) and other factors (\textit{knowledge, verification tendency, and worthiness}), highlighting a multifaceted set of determinants.


To deepen our understanding of these matters, particularly concerning RQ3, we formulated RQ4 and RQ5. To answer the subsequent RQs, we conducted \textbf{Study 2}, described in detail in \S\ref{sec:study-two}. Study 2 consisted of semi-structured interviews with 12 participants chosen from the 220 survey respondents. These interviews not only supplemented our survey insights but also facilitated a more nuanced exploration of the subsequent research questions from a moral perspective.

\begin{description}
    \item [RQ4] How do users' experiences with deceptive responses from ChatGPT influence their trust and reliance on the technology, and what methods do they employ to manage these situations?
    \item [RQ5] What are users' perspectives on the need for regulatory measures and improvements for ChatGPT, and who do they believe should be held responsible for managing the risks associated with deceptive responses?
\end{description}

\textbf{RQ4-5 Highlights:} Study 2 revealed nuanced insights into users' mental models and experiences with ChatGPT, emphasizing a blend of daily and professional utilizations. Participants reported a generally positive outlook on ChatGPT's conversational capabilities, highlighting its efficiency and utility over traditional tools, yet also expressed concerns over its potential for deception and the ethical implications of its use. Specifically, when encountering deceptive information, there seems to be a notable shift in users' trust levels and attitudes towards ChatGPT. Initially, some participants displayed a low trust level, which either increased upon recognizing ChatGPT's advantages or decreased after participants noticed inaccuracies. This led participants to take a more cautious approach when using the technology, which indicates the important role that accuracy, reliability, and explanatory transparency play in shaping user trust. 

\textbf{General Highlights:} Our findings from Study 1 and 2 emphasise the complex dynamics of responsibility for ChatGPT's deceptive outputs, with participants attributing responsibility to developers, hosting platforms, and, to a lesser extent, users of the technology. The results also indicate a consensus on the need for enhanced verification strategies, user education, and regulatory frameworks aimed at mitigating the risks associated with deceptive information. Finally, our study's results highlight the need of addressing ethical standpoints in the development and use of AI technologies like ChatGPT, advocating for a balanced approach that considers user empowerment, technological improvements, and robust safeguard strategies to enhance trustworthiness and mitigate potential harms.



The paper is structured as follows: In \S\ref{sec:study-one} we describe the method and describe the results from Study 1. In Section \ref{sec:study-two}, we describe the method and results from Study 2. Then, in \S\ref{sec:discussion}, we discuss, integrate, and summarise the overall results and insights from both studies. After that,  \S\ref{sec:rel_work} presents the related work in the area of Deceptive AI \& Society and contextualize our approach within this area of research. Finally, in \S\ref{sec:future_dir} we discuss future directions in communicative AI agent technologies, and in  \S\ref{sec:conclusion} we conclude the paper.

The supplementary material, including anonymized user data, quantitative analysis, interview scripts, and a detailed description of themes and codes, is also available in the anonymised OSF repository\footnote{Link anonimised for review: \url{https://osf.io/uf5v3/?view_only=da0b14aaabe34bca811f85e0e5f65882}} for this study.


\section{Study 1: Online Survey}
\label{sec:study-one}

This study addresses questions R1-3, focusing on an in-depth examination of user opinions concerning deceptive behaviours exhibited by ChatGPT. More precisely, Study 1 collects user responses and insights about their experiences with ChatGPT, with a particular emphasis on identifying instances of deceptive behaviour encountered during interactions, such as types of deceptions and the specific domains where these behaviours commonly occur.


\subsection{Method}
In this section, we conducted a survey study with 220 participants from both UK an US. We provide an overview of the data collection and Chi-Square tests and post-hoc analysis methodology. 
This study was reviewed and approved by our institution's IRB.

\subsubsection{\textbf{Participants}} We recruited participants via Prolific\footnote{\url{https://www.prolific.co/}}. Using a screening survey, we selected 220 participants who met the following criteria: (a) engagement with ChatGPT in the past six months, (b) experience with deceptive responses during their interactions, and (c) residence in either the UK or USA. Choosing participants from the US and the UK for the study on ChatGPT's deception responses is justified by their high English proficiency and significant digital literacy, which ensure accurate engagement with AI. These countries' advanced technology adoption and established regulatory frameworks provide a pertinent backdrop for exploring AI interactions and user expectations, offering a comprehensive view on the impact of deceptive AI responses within a Western context. 
To ensure  data quality, we recruited \emph{high-reputation} participants with at least 100 submissions and an approval rate of 95\% or more on the Prolific recruitment platform~\citep{peer2014reputation,such2017photo}. We obtained valid data from 220 participants. See participant demographics summarised in Table \ref{tab:demographics}.
\begin{table}[h]
    \centering
    \caption{Demographics of survey participants.}
    \small
    \begin{tabular}{llc}
         \toprule
         && \#Participants \\
         \midrule
        \multirow{3}{*}{Gender}&Female & 108\\
        &Male & 110\\
        &Prefer not to answer & 2\\
        \midrule
        \multirow{6}{*}{Age}& 18-24&41 \\
         &25-34 & 86\\
         &35-44 & 53\\
         &45-54 & 27\\
         &55-64 & 8\\
         &65+ & 5\\
            \midrule
        \multirow{6}{*}{Employment Details}& Full-time employment& 130\\
        &Full-time student & 12 \\
        &Part-time employment/student & 35\\
        &Not employed, job seeking&  17\\
        &Not employed, not seeking& 14\\
        &Others&12\\
        \midrule
     \multirow{6}{*}{Education}& Middle School& 1\\
     &High school & 31\\
     &Sixth-form college/school & 41\\
    &HND; or University & 94\\
    &Postgraduate school&40\\
    &Doctorate&12\\
    &Prefer not to answer&1\\
    \midrule
    \multirow{3}{*}{Income}& Low& 107\\
    &Middle&79\\
    &High&27\\
    &Prefer not to answer&7\\
        \bottomrule
        Total &&220\\
    \end{tabular}
    \label{tab:demographics}
\end{table}

\subsubsection{\textbf{Instrumentation \& Procedure.}}
Our survey was created and hosted on Qualtrics\footnote{\url{https://www.qualtrics.com/}}. Initially, participants received an information sheet explaining the purpose of the study, the nature of participation, and confidentiality assurances. This was followed by a consent form, which participants completed to confirm their willingness to participate in the study. The main body of the survey 
focusing on participants' general use of ChatGPT, their encounters with deceptive responses, and their views on ChatGPT's worthiness, among other aspects. The survey concluded with questions regarding participants' demographics and their willingness to be considered for a follow-up in-depth interview. On average, the survey took about 4 minutes to complete, and we compensated participants at £16 per hour. 
Two pilot studies (N=2 each) were conducted to refine our survey instrument. These studies focused on assessing question clarity, layout understanding, and survey logic effectiveness. Feedback from these pilots led to adjustments in question wording and survey design. Data from the pilot studies were used solely for refinement purposes and excluded from the final analysis.

\subsubsection{\textbf{Data Analysis.}}
We first conducted descriptive statistics to address RQ1. This was followed by employing a 
Chi-square test~\cite{mchugh2013chi} to investigate whether and how demographics and personal factors - encompassing users' knowledge of LLMs, frequency of ChatGPT usage, frequency of receiving deceptive responses, and the frequency of verifying ChatGPT's responses - influence perceptions of ChatGPT's worthiness, responsibility, and users' post-behavior.


\begin{figure}[t]
    \centering
     \adjustbox{trim={0.0\width} {0.0\height} {0.0\width} {0.0\height},clip,center}{
    \includegraphics[width = 0.7\textwidth]{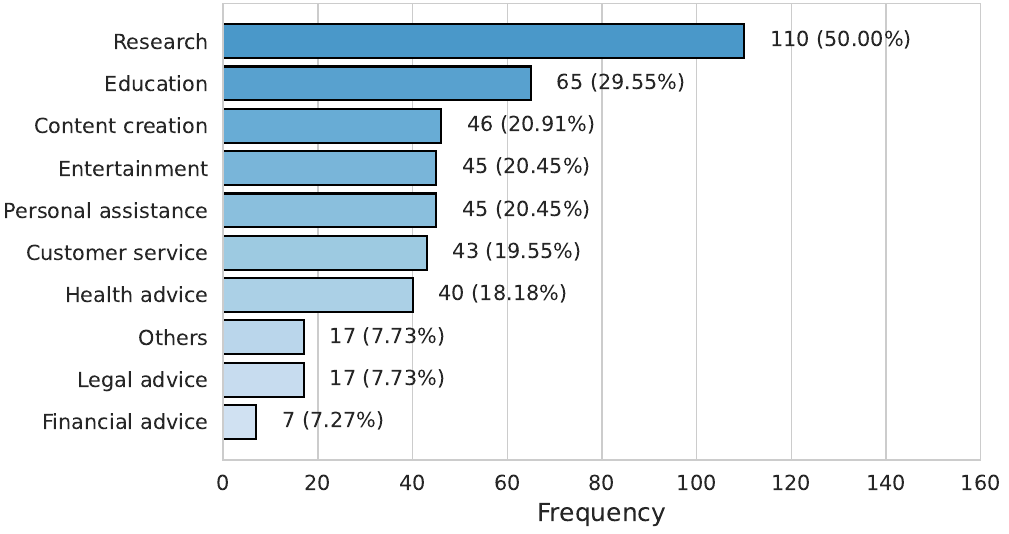}}
    \caption{Common Contexts for Deceptive Behaviour}
    \label{fig:deception-contexts}
\end{figure}

\subsection{Study 1 Results}




\begin{figure}
    \centering
     \adjustbox{trim={0.0\width} {0.0\height} {0.0\width} {0.0\height},clip,center}{
    \includegraphics[width = 0.7\textwidth]{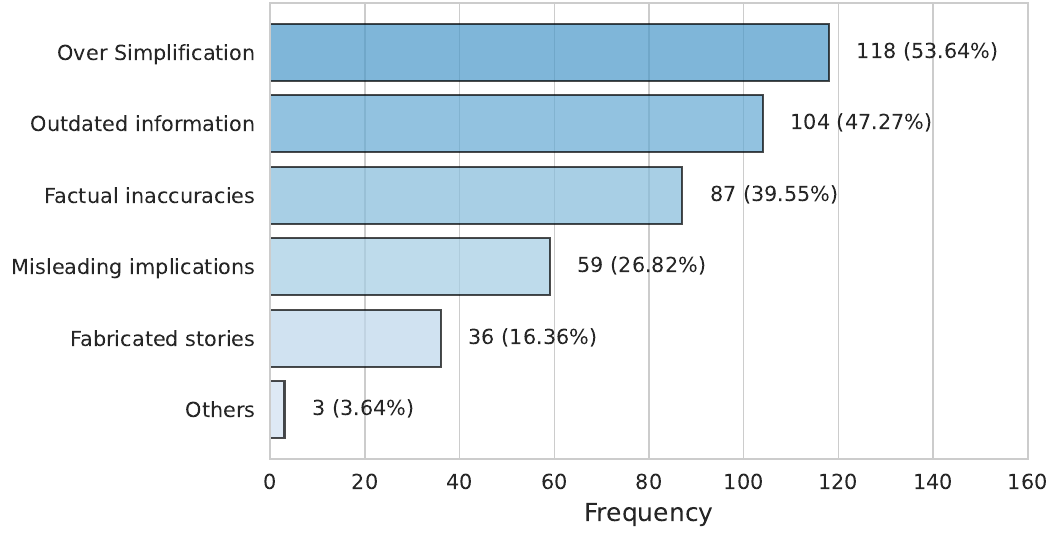}}
    \caption{Common Forms of Deceptive Behaviour}
    \label{fig:deception-domain-type}
\end{figure}


\subsubsection{\textbf{Common Forms and Contexts of Deception.}} \label{sec:common-form-context-of-deception}
We illustrate the common deceptive categories in Figure \ref{fig:deception-domain-type}, including ``Oversimplified", ``Outdated", ``Inaccurate", ``Misleading", and ``Fabricated". We observed that the most frequent form of deceptive response encountered by ChatGPT users is \emph{Overly Simplified Answers}, reported by 53.64\% of participants. This category is \textbf{the only one} surpassing the 50\% threshold. This implies that a substantial amount of the information may not provide a thorough comprehension of the subject topic, potentially leading to misunderstandings or misinterpretations.

\begin{figure}[!h]
    \centering
    \includegraphics[width=\linewidth]{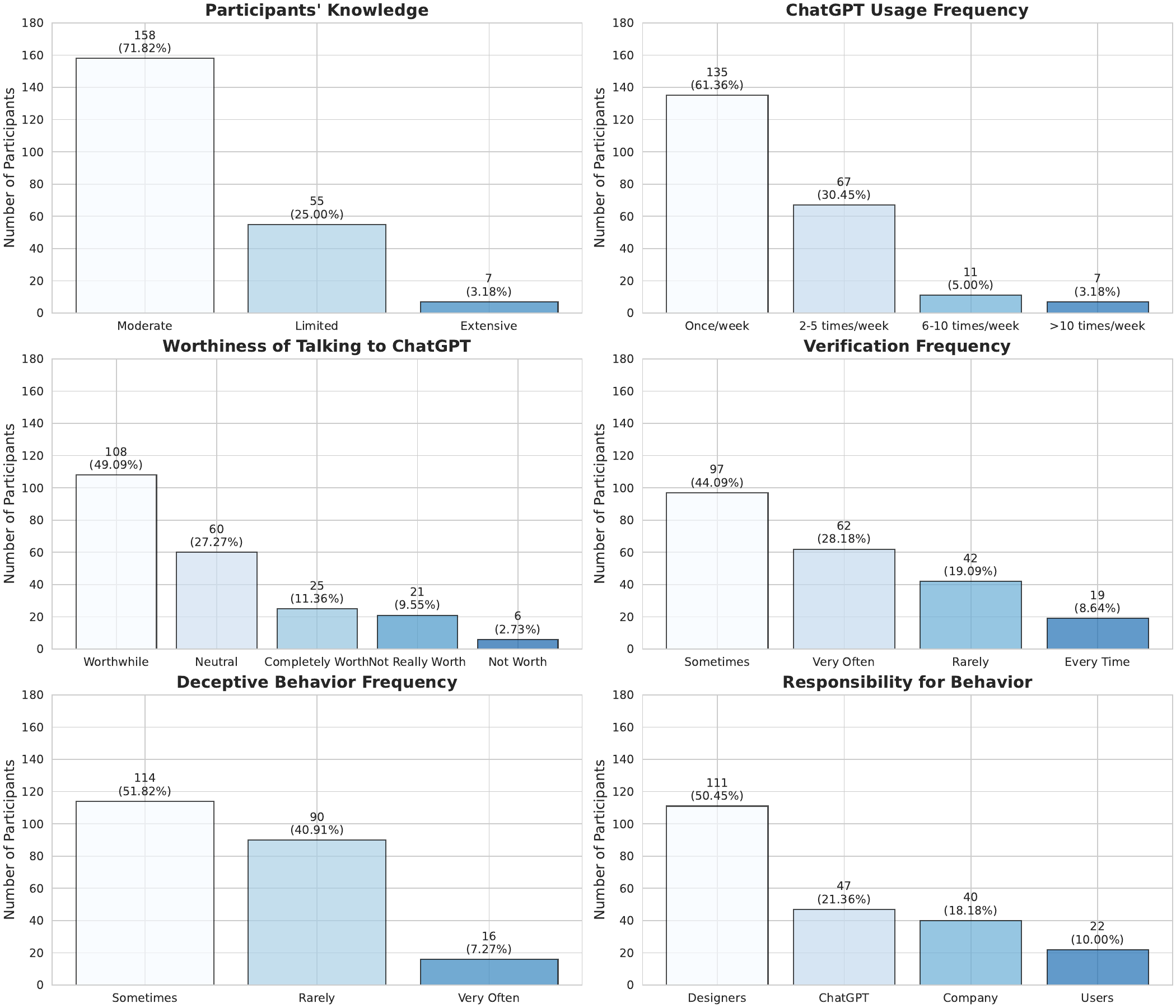}
    \caption{Descriptive analysis of survey responses regarding: (1) Participants' Knowledge, (2) ChatGPT Usage Frequency, (3) Worthiness of Talking to ChatGPT, (4) Verification Frequency, (5) Deceptive Behavior Frequency, and (6) Responsibility for Behavior.}
    \label{fig:decriptive-analysis}
\end{figure}

Figure \ref{fig:deception-contexts} shows the common areas that encounter deceptive behaviours in ChatGPT, including education, entertainment, personal assistance, customer service, health, financial advice, and other legal advice. It indicates that a significant proportion of participants, about half, experienced deceptive responses while discussing \emph{research-related} topics with ChatGPT. Furthermore, \emph{Education} is affected to be the second most common context (29.55\%) for receiving deceptive information.


\subsubsection{\textbf{Descriptive Analysis of Personal Factors}}\label{sec:personal-factor-descriptive}

Regarding their knowledge of AI, 71.82\% of participants rated their knowledge as \textit{moderate}, and 3.18\% considered themselves experts. Moreover, the majority of users do not frequently use ChatGPT, with over half (61.36\%) utilizing it just \textit{once a week}. The second-largest group of participants (30.45\%) uses it between \textit{2-5 times a week}. Only 5\% of users engage with it \textit{6-10 times} weekly, while a mere 3.18\% do so \textit{more than 10 times a week}. When it comes to the frequency of using ChatGPT, a majority of participants, constituting 51.82\%, reported that they \textit{sometimes} receive deceptive information, whereas 40.91\% of participants indicated that they \textit{rarely} encountered deceptive information. A minimal fraction of the participants, approximately 7.27\%, stated that deceptive behaviors appeared \textit{very frequently}. In terms of verifying ChatGPT's responses, 44.09\% reported doing so \textit{occasionally}, while 8.64\% \textit{always} checked the accuracy of the information provided. 

\subsubsection{\textbf{User's Opinions on ChatGPT's Worthiness, Responsibility for Deception, and Their Behavior Change}} \label{sec: descriptive-four-main-factors}

In evaluating the worthiness of engaging with ChatGPT, we utilized a scale from 1 to 5, where 1 signifies the lowest value, indicating it is not worth engaging at all, and 5 indicates the highest value, suggesting it is entirely worthwhile. The majority of respondents, over 60.45\%, rated their experience \textit{positively}. Only a small fraction, 2.73\%, felt it was not worth engaging with, giving it the lowest rating of 1. 

We were also keen to investigate whether users would continue to use ChatGPT after receiving incorrect answers. Surprisingly, we found that after experiencing deceptive behaviour, 54.1\% of participants continued to use ChatGPT, 38.6\% chose to reduce their usage, and only 7.3\% decided to stop using it altogether.

In terms of the question related to responsibility for ChatGPT's potentially misleading outputs, a majority of 50.45\% pointed to its \emph{designers or creators}. Approximately 21.36\% attributed the responsibility to \emph{ChatGPT itself}, while only 10\% believed that users should be accountable.  

\subsubsection{\textbf{The impact of participants' demographics and personal factors on Worthiness, Responsibility, and Behavioral Response.}}

\begin{table}[tbh]
    \centering
    \caption{Results of the Chi-square test (where `-' indicates non-applicability, and empty cells represent non-significant outcomes, hence not included in this table).}
    \scalebox{0.9}{
    \begin{tabular}{l|ccc|ccc|ccc}
    \toprule
\textbf{Factors}& \multicolumn{3}{c|}{\textbf{Worthiness}} &\multicolumn{3}{c|}{\textbf{Responsibility}}&\multicolumn{3}{c}{\textbf{Behavioral Response}}\\

& $\chi^2$ & $p$ &  Sig.? &$\chi^2$ &$p$&Sig.?&$\chi^2$ &$p$&Sig.?\\
\midrule
\textbf{Gender}&&&&&&&10.090&0.006&Y\\
\textbf{Age}&&&&&&&25.140&0.005&Y\\
\textbf{Employment}&&&&&&&&&\\
\textbf{Education}&&&&25.925&0.039&Y&&&\\
\textbf{Income}&&&&&&&&&\\
\midrule
\textbf{Knowledge}&&&&&&&14.636&0.006&Y\\
\textbf{Deception Fre.}&22.084&0.005&Y&19.355&0.004&Y&&&\\
\textbf{Use Fre.}&34.208&$<$0.001&Y&&&&-&-&-\\
\textbf{Verification}&-&-&-&&&&20.826&0.002&Y\\
\textbf{Worthiness}&-&-&-&-&-&-&98.514&$<$0.001&Y\\

\bottomrule
    \end{tabular}}
    \label{tab:chi-square}
\end{table}


As demonstrated in Table~\ref{tab:chi-square}, the differences of \emph{Behavioral Response} among different \textit{genders} are evident in the comparative percentages. Males predominantly chose to keep using ChatGPT, while a significantly higher percentage of females (47.92\%) opted to reduce its usage. As for various age groups, those aged 18-24 demonstrate a significant higher likelihood of keeping use ChatGPT in comparison to other age brackets. Moreover, participants with \textit{postgraduate degrees} are significantly more inclined to attribute ChatGPT's deceptions to \textit{itself}, whereas those with \textit{middle school} and \textit{university} education levels tend to hold the \textit{company} responsible. Conversely, \textit{high school} and \textit{doctoral} degree holders more often assign responsibility to the \textit{designers}. These findings represent relative outcomes across different educational levels.

Regarding the impacts caused by personal factors, the \textit{frequency of encountering deceptive responses} significantly affects participants' perceived \textit{worthiness} of ChatGPT and their views on \textit{responsibility}. Notably, individuals who receive deceptive responses \textit{very often} tend to view ChatGPT as \textit{slightly unworthy} of engagement and are more likely to attribute responsibility for these deceptions to the \textit{designers}. In contrast, those who \textit{rarely} encounter such responses tend to consider it \textit{slightly worth} engaging with and are inclined to hold the \textit{company} accountable. These findings align with general expectations and common understanding. The \textit{frequency of using} ChatGPT is significantly correlated with users' perception of its \textit{worthiness}. However, the most frequent participants (over 10 times a week) are inclined to view ChatGPT as ``slightly not worth talking to". Conversely, those using it 6-10 times a week lean towards seeing it as ``slightly worth talking to". The least frequent users (once a week) predominantly opt for a \textit{neutral} stance in their assessment. For factors that significantly affect participantss post-behaviour, individuals possessing \textit{extensive and expert knowledge} of LLMs show a significant preference for \textit{keeping} their use of ChatGPT. In contrast, those with \textit{lower} level knowledge are notably more inclined to \textit{reduce} their usage. Individuals who \textit{sometimes} verify responses generated by ChatGPT \textit{keep} using the service, while those that verify their responses \textit{every time} are inclined to \textit{reduce} their usage.

\section{Study 2: Semi-structured Interview Study}
\label{sec:study-two}
While Study 1 focuses on answering research questions RQ1-3 in a descriptive manner, we aim to explore further how users perceive their encounters with ChatGPT's deceptive behaviours and how these experiences impact their usage, trust, and future preferences (RQ4 and RQ5). Study 2 provided insights into users' mental models and experiences with ChatGPT, showing its use for both daily and professional purposes.


\subsection{Method}
In this section, we conducted semi-structured interviews with 12 participants from both the UK and the US. We provide an overview of the data collection using thematic analysis. 
This study was reviewed and approved by our institution's IRB.

\subsubsection{\textbf{Participants.}}
From the original pool of 220 survey respondents, 14 participants (12 for final analysis and 2 for pilots) were chosen for the in-depth interviews. This approach facilitated seamless progression to more detailed explorations during the interviews, leveraging the participants' pre-established familiarity with the survey themes. Recruiting interviewees from our survey respondents not only streamlined the research process by eliminating the need for a new recruitment phase but also minimized potential sample bias. Selection was based on their willingness to participate further, and a stratified sampling method was employed to balance the participants' variances, especially in significant factors shown in Table~\ref{tab:chi-square}. For instance, we found that participants' \textit{age} and \textit{gender} significantly influence their perceptions of post-behavior. To comprehensively investigate this phenomenon, we recruited participants who exhibited a wide range of age groups and genders in our survey study. The demographics of these participants are summarised in Table~\ref{tab: demo-interview}.

\begin{table}[!h]
\centering
\caption{Demographics of interview participants. Note `-' indicates that this participant had retired before the release of ChatGPT and therefore was not eligible to use ChatGPT at professional work.}
\label{tab: demo-interview}
 \scalebox{0.8}{
\begin{tabular}{lllllcc}
\toprule
\multicolumn{1}{c}{\multirow{2}{*}{Par.}} & \multicolumn{1}{c}{\multirow{2}{*}{Age}} & \multicolumn{1}{c}{\multirow{2}{*}{Gender}} & \multicolumn{1}{c}{\multirow{2}{*}{Knowledge}} & \multicolumn{1}{c}{\multirow{2}{*}{Job}} & \multicolumn{2}{c}{Usage}                                         \\ \cline{6-7} 
\multicolumn{1}{c}{}                     & \multicolumn{1}{c}{}                     & \multicolumn{1}{c}{}                        & \multicolumn{1}{c}{}                           & \multicolumn{1}{c}{}                     & \multicolumn{1}{c}{Professional work} & \multicolumn{1}{c}{Daily life} \\ 
\midrule
P1  & 18-24  & Male& Extensive& NLP researcher& \checkmark & \checkmark \\
P2  & 25-34  & Female& Extensive& University lecturer in Information& \checkmark& \checkmark \\
P3& 45-54&Female&Limited&Chef& $\times$ &\checkmark \\
P4&65+&Male&Moderate&Retired&-&\checkmark \\
P5&35-44&Male&Moderate&Solicitor&\checkmark &\checkmark \\
P6&25-34&Male&Moderate&Clinical researcher&\checkmark&\checkmark \\
P7&25-34&Male&Extensive&Robotics Company Engineer&\checkmark&\checkmark \\
P8& 25-34&Female&Moderate&Marketing manager &\checkmark&\checkmark \\
P9& 45-54&Male&Limited&University professor in Music theory&\checkmark&\checkmark \\
P10&35-44 &Male&Limited&Civil servant&\checkmark&\checkmark \\
P11&18-24 &Female&Moderate&Software tester&$\times$ &\checkmark \\
P12&25-34 &Female&Moderate&Student&\checkmark&\checkmark \\
\bottomrule
\end{tabular}
}
\end{table}


Potential participants were approached through the Prolific platform. 
All interviews were conducted virtually via Zoom\footnote{\url{https://zoom.us/}}, each averaging approximately 30 minutes in duration. With prior consent from the participants, all sessions were audio-recorded. 
Each participant was compensated at £20 per hour through Prolific. The interviews were conducted from from January to February 2024. 

Among the 12 participants interviewed, there were 7 (58.3\%) males and 5 (41.7\%) females, with their ages spanning almost all age groups: 2 (16.7\%) individuals aged 18-24, 5 (41.7\%) individuals aged 25-34, 2 (16.7\%) individuals aged 35-44, 2 (16.7\%) individuals aged 45-54, and 1 (8.3\%) individual aged 65 or above. Among these participants, 3 (25\%) possess extensive knowledge of ChatGPT, 6 (50\%) have moderate knowledge, and 3 (25\%) have only limited knowledge.

\subsubsection{\textbf{Interview Protocol.}}
The interview protocol was semi-structured, known for enabling detailed and comparable qualitative data~\citep{knott2022interviews}. 
During each interview, researchers strategically posed opportunistic follow-up questions as necessary, aiming to comprehensively capture the participants' experiences.  The interview script contains questions on the following topics:
\begin{enumerate}
    \item 
    Participants' usage of ChatGPT in both personal and professional settings, their integration of its assistance with their own skills, and their motivations for using it. It examines specific tasks where ChatGPT is crucial, their knowledge of its capabilities, and comparisons with other tools like search engines. 
    \item Participants' experience with deceptive or inaccurate responses from ChatGPT, their methods of handling such situations, and the impact on their perceptions of ChatGPT's reliability and subsequent behaviour.
    \item Factors affecting trust in ChatGPT and participants' views on its future reliability. This includes responsibility for deceptive responses and potential risks of over-reliance, especially for vulnerable groups.
    \item Participants' preferences and expectations for regulatory measures and improvements to ChatGPT.

\end{enumerate}

To refine the protocol, we initiated the process with a pilot study involving two participants following the same selection procedure. This enabled us to refine the interview structure, ensuring each question was clear, understandable, and effectively designed to elicit the targeted information. The pilot interviews were excluded from the final dataset for analysis. 

\subsubsection{\textbf{Data Analysis.}}
We performed a thematic analysis~\citep{braun2012thematic} to process participant responses. We started with two researchers independently coding the first interview transcript to identify salient codes, thereby establishing an initial codebook. This preliminary codebook was then collaboratively refined during the analysis of the second interview transcript, where the researchers engaged in a detailed discussion to reach consensus on the codes that was applied to the remainder of the interview transcripts. Then, the two researchers independently coded subsequent transcripts, when new codes emerged, the researchers met to discuss these new findings and, where necessary, made amendments to the codebook. This iterative process continued until no new codes were identified, indicating a point of code saturation, which occurred after analyzing seven transcripts. Upon completing the coding of all interviews, the researchers collectively reviewed and deliberated on the potential themes. This collaborative review process was instrumental in ensuring the thematic saturation and in achieving a consensus on the final themes. To ensure a high level of inter-coder reliability throughout the study, Cohen's kappa statistics~\citep{fleiss2013statistical} were computed for each interview transcript, and the final average is .84, indicating \textit{substantial agreement} between the researchers. This suggests that the coding scheme was applied consistently, lending credibility to the thematic analysis conducted. 
The finalised codebook and themes can be found in the OSF repository\footnote{Link anonymised for review. \url{https://osf.io/uf5v3/?view_only=da0b14aaabe34bca811f85e0e5f65882}}. 


\subsection{Study 2 Results}
The qualitative findings are reported below. We have edited the reported quotes to remove filler words e.g., \textit{``umm", ``like", ``ah-ha"}) with \textit{Hemingway Editor\footnote{\url{https://hemingwayapp.com/}}} used to indicate where quotes have been condensed for brevity.

\subsubsection{\textbf{Mental Models of \& Experiences with ChatGPT}} 
\label{sec:mental-model}
This section reported users' interactions with ChatGPT, focusing on their uses, motivations, and general attitudes.
\paragraph{Daily vs. Professional Purposes.} During the interview, one of the questions we asked was about the purpose of using ChatGPT, and participants mentioned the capabilities of ChatGPT. Not only do they use it in their daily life but also as part of their job or profession.  All 12 participants have been integrating ChatGPT into their everyday activities, with 9 (75\%) of them also employing it for professional purposes. ``Daily Purpose" includes Artwork Creation, Writing Help, Idea Generation, and Social Interaction, which are typically used for personal enjoyment and everyday tasks.  On the other hand, ``Professional Purpose" encompasses Academic Content Creation, Newsletter Generation, Automated Grading, Medical QA, Demonstrations, Legal Advice, Information Integration, Code Generation, and Assistant Tools, which are primarily intended for specialized tasks in professional settings such as education, healthcare, legal, and technical fields. 
Specifically, we find that professional use often involves highly specialized technologies to optimize the ChatGPT. For example, P6 (Clinical Researcher) mentioned their institution's customized version of ChatGPT. Similarly, P1 (NLP researcher) demonstrated a uniquely comprehensive use of ChatGPT and further extended to experimenting with various other LLMs, while the others used plain versions of GPTs. Then, in discussing the primary motivations for utilizing ChatGPT, it was observed that aside from P1, the other participants' interest was attributed either to an intrinsic curiosity about the capabilities of ChatGPT or to conformity with prevailing social trends (or `Herd mentality'). 

\paragraph{Positive Perspectives.} Participants in the study underscored ChatGPT's conversational effectiveness, efficiency, and overall utility. They particularly noted its superiority over traditional tools, such as search engines and translation tools, in conducting similar tasks. Furthermore, the adoption of ChatGPT by various academic and companies was highlighted. For example, P2 remarked on the potential or existing inclination of universities to incorporate generative AI technologies like ChatGPT, plus P6 stated \textit{``Our company has also created its own version of ChatGPT. We have a dedicated team responsible for its development and maintenance. Employees are taught and encouraged to use it for work-related tasks, like finding information about a specific drug, [...]"}. Finally, participants also expressed a belief that ChatGPT will be improved in the future and that deception issues will eventually be resolved.

\paragraph{Negative Perspectives.} Compared to positive attitudes, there is a predominant emphasis on negative perspectives expressed by participants during the interviews. This includes 1) inherent negative perceptions regarding AI technology, 2) negative consequences of using ChatGPT that have already occurred and concerns about potential future ones. Note that we separate out the negative consequences that ChatGPT's deceptive responses cause or may cause and discuss them in \S\ref{sec: deception-and-reaction}, and 3) general feelings that current ChatGPT is \textit{``limited in its capabilities"} [P6, P7, P12] or \textit{``not ready yet"} [P2]. Among them, we observed that there is a concern among participants regarding the privacy and safety implications of using ChatGPT in general [P1, P2, P6, P12]. Participants legitimately question whether their personal data could be compromised when posing queries of a personal nature, as well as whether the integrity of information sourced through ChatGPT is assured, particularly in instances where it aggregates data from various origins. 

\paragraph{Knowledge of ChatGPT.} Participants exhibited varying levels of understanding and assumptions about ChatGPT's operations, ranging from its source of information to its capabilities and limitations. Interestingly, despite the varying levels of self-reported knowledge among users (see Table\ref{tab: demo-interview}), their understanding of ChatGPT's operations, as discussed during the interviews, was remarkably consistent. Several participants [P6, P7, P8, P9] described ChatGPT as a data mining or scraping tool, leveraging large repositories of internet data, including a mix of pre-existing knowledge and generative capabilities based on predictive algorithms. There was a recognition of the vast amount of data ChatGPT has access to, including digitized books and potentially internet forums, although there was uncertainty about its access to subscription-based journals and books. Only one participant [P3] believed that ChatGPT’s information source was based solely on user input.

\subsubsection{\textbf{Deception and User Reactions}} In this section, we summarize user experiences with deceptive information from ChatGPT, focusing on their reactions and perceptions of these encounters. 
\label{sec: deception-and-reaction}

\paragraph{Deceptive Information Received by Participants.} Most of the deceptive information generated by ChatGPT mentioned by participants is consistent with those shown in Figure \ref{fig:deception-domain-type}. In addition, ChatGPT also performs poorly on other tasks, such as their inability to mimic the writing style of specific individuals [P4]. Interestingly, misleading responses generated by ChatGPT are able to manifest themselves both explicitly and implicitly to our participants. For instance, claims made by ChatGPT like \textit{``Vaccines often cause more harm than good"} and \textit{``You can always trust news shared on social media"} are obviously dubious. Whereas what participants find particularly troublesome is that ChatGPT sometimes \textit{``produces partially correct information or advice, making it extremely difficult to discern truth from falsehood''} [P1, P6-12]. A very interesting example experienced by P8:

\begin{displayquote}

 \textit{[...] asked ChatGPT to generate a list of current thought leaders in the marketing industry. And it listed some names and I don't remember who they are now, but some names were real, some were just like some totally made-up person that wasn't a real thought leader in the marketing industry.
}
\end{displayquote}

\paragraph{Response Checking and Behavior Changing.} With regard to the deceptive information provided by ChatGPT, essentially all participants were aware of this problem, with only P5 mentioning that they would perform a detailed check on almost every response given by ChatGPT. P2 indicated that the proactive verification behaviour only began after the first time noticed deceptive information in ChatGPT’s response, wherein P2 realised \textit{``indeed ChatGPT can also make mistakes, okay, let me check these solutions to make sure it's reliable."} Most commonly, participants [P1, P2, P4, P6, P8-12] (9/11) decided whether to conduct detailed fact-checking based on the importance of the context in which they want ChatGPT to function. For example, P10 mentioned \textit{``it depends on what you're using it for, I was using it for something like a medical diagnosis like something critical, I think that we want to be checking things, [...]".} Moreover, a significant obstacle arises when participants pose questions to ChatGPT that fall outside their knowledge or expertise, making it difficult for them to verify responses and identify potential inaccuracies [P5, P7]. P5 even described feeling \textit{``incompetent"} when addressing unfamiliar responses and candidly admitted, \textit{``I would take [ChatGPT's response] as the truth."} This predicament is particularly pronounced among individuals with an inherent inclination to trust, who may lack the motivation to scrutinize the responses further, thus unwittingly accept inaccurate information.

After experiencing deceptive responses from ChatGPT, participants' subsequent behaviours diverge. 8 participants [P2-4, P7-9, P11, P12] reduced their usage of ChatGPT for the given task.  As for the reason, it includes \textit{``I've kind of gone back to Google after an initial enthusiasm"} [P4] and realises this doesn't help and \textit{``might cause messy for serious tasks"} than just for entertainment [P2, P8]. However, the remaining 3 participants overlooked the deceptive responses and maintained their original frequency of use. They justified this decision by explaining that they relied on ChatGPT for very simple tasks, where deceptions were readily identifiable.

\paragraph{Correct Actions and Efficacy of ChatGPT Responses.} Participants employed diverse strategies to rectify ChatGPT's responses. Some opted for a straightforward correction by responding with a brief assertion such as \textit{``you are wrong''} without furnishing additional instructions [P3, P5, P8]. Conversely, others also offered guidance or specific requirements in their prompts [P1-4, P7, P8, P10-12], for instance, P4 mentioned \textit{``[...], I was feeding it previous examples of articles I'd written, [...]"} and \textit{``[...] say, this code didn't run, you know, I got this error message, and I put it back to [...]"}. Subsequently, it was observed that a greater number of participants (n = 5) did not experience an improvement in ChatGPT's responses, whereas only 4 participants reported an improvement. It is noteworthy that these improvements were all predicated on participants providing explicit instruction. Participants also specifically mentioned the limitations of ChatGPT in functionalities (modalities) extending beyond text, such as generating images [P1, P3, P4, P12]. When participants provided additional guidance to ChatGPT, they observed that the output \textit{``got incrementally worse"}. To elucidate further, ChatGPT even induced a feeling among users, described as \textit{``It seemed to get confused with further instructions, [...], every time I added something, it just got more murky and, it lost the integrity of what it was trying to be, [...]"}[P3]. P10 stands out as the only individual who does not attempt to correct ChatGPT upon discovering deceptive information, demonstrating indifference towards its accuracy with a remark, \textit{``well, that's wrong. That's kind of the end of it.”}

\paragraph{Perceptions on Why ChatGPT Generates Deceptive Responses} But when it comes to the reasons behind the deceptive information, only P1, P5, P7, P8, P11, and P12 offered their speculations. The other participants expressed that they found it strange but were unsure of the specific causes. P5, P8, and P11 noted that their impression of ChatGPT is that its objective appears to be to strive or attempt to provide responses that seem logically coherent or to assemble elements that sound correct, regardless of the question or request, even though the actual information provided may be inaccurate. P7 posited that ChatGPT may occasionally misinterpret specific words or sequences of words within user prompts. Finally, P12 mentions that maybe the information in the ChatGPT knowledge base or training data is inherently wrong.

\paragraph{Negative Consequences Regarding Deceptive Information} Fortunately, to date, none of the 12 participants have faced any risks or encountered serious consequences as a result of receiving deceptive information. However, most have reported awareness of individuals within their networks who have experienced such issues, or they have expressed concerns regarding the potential harm deceptive information may cause.


Participants [P1, P5, P6, P8-11] expressed concerns about the consequences of specific groups receiving deceptive information from ChatGPT. 
To provide a more details, we identify the following demographics that participants have deemed particularly susceptible to the deceptive information disseminated by ChatGPT.
\begin{itemize}
    \item Kids and the Elderly [P5, P6, P9]. In participants' minds, kids and the elderly are particularly vulnerable due to a combination of developmental, cognitive and technological factors. Kids are still developing critical thinking skills and inherently more trusting, while the elderly, might be affected because of the potential sensory and cognitive declines. For instance, P6 thought kids \textit{``do generally trust people. And they're not as sceptical as adults, [...]."}
    \item (Young) students [P6, P8-11]. Young students are particularly vulnerable to deceptive information from AI technologies like ChatGPT due to their specific need for quick answers, and their familiarity and comfort with using such technologies. This demographic's tendency to rely heavily on AI for academic assistance, without adequately verifying the information's accuracy or engaging deeply with the material, heightens their risk of being misled. P9 specially mentioned, that young students dependence on ChatGPT for quick solutions can atrophy their ability to independently evaluate arguments, and P9 articulated apprehensions regarding the long-term effects of such dependence, positing that \textit{``I suspect the more that we rely on ChatGPT, it's possible that our own skills especially the critical thinking ability will diminish, [...]"}
    \item Individuals Unable to Afford Healthcare Services [P8]. The participant believes that due to high healthcare costs in the US, people who rarely visit doctors or cannot afford them might use ChatGPT for initial health advice. They contrast this with the UK\footnote{Hospital treatment is free to people who are ``ordinarily resident" in the UK. \url{https://www.jpaget.nhs.uk/patients-visitors/overseas-visitors/information-for-people-seeking-free-nhs-hospital-treatment/}}, suggesting the issue might be less severe there, and indicate this behaviour is a result of financial barriers to accessing medical care.
    \item Non-tech-savvy [P1, 10, P11]. People with no technical background are likely to overtrust AI due to unfamiliarity with technology's limitations and a belief in its infallibility, as highlighted by P1's observation. They might assume the computer's infallibility, believing P10, ``\textit{the computer must know; it knows all this other stuff, so it must be right."} This naivety can lead to uncritical acceptance of potentially inaccurate information as noted by P1 and P11. Additionally, P11 mentioned that individuals with mental health issues are particularly vulnerable, as they may find it even more challenging to discern the reliability of information provided by AI.
 
\end{itemize}


\subsubsection{\textbf{Responsibility and Trust}}
\label{sec: trust-and-responsibility}
By asking specific questions, we delve into the complex dynamics of responsibility and trustworthiness concerning ChatGPT’s deceptive information. In terms of the perceived responsibilities of various stakeholders, we distinguish between `developers', defined as employees responsible for technical product development, and the `hosting platform', the entity or company managing and operating the product. In the context of ChatGPT, `developers' refers to the engineers, while OpenAI serves as the `hosting platform' overseeing its deployment and API management.

\paragraph{Unintentional Deception and No One Should be Responsible} When questioning responsibility for ChatGPT's deceptive behavior,  several participants [P1, 3-5, P6, P7, P9, P11, P12] (9/12) emphasized that ChatGPT, lacking `consciousness', does not engage in deception intentionally. And, as such, should not be held responsible for such actions. While acknowledging the \textit{``existence of AI systems intentionally trained by humans to disseminate misleading information, which could be deemed deceptive,"} P5 still believed ChatGPT should not be categorized like that. Furthermore, as described by P7, 
\textit{“[…] It hasn't lied to me. I take it as incompetence. I take it as lack of knowledge, [...]"}, highlighting it as a limitation of the AI technology behind ChatGPT.

\paragraph{Developers' Responsibility}
[P1, P2, P5, P7, P9, P10-12] (8/12) participants attributed the responsibility for ChatGPT's potentially deceptive outputs to its developers, 
emphasizing the developers' crucial role in designing algorithms that are discerning in data verification and source selection from the outset. As creators of ChatGPT, developers are tasked with safeguarding the integrity and accuracy of the content provided, and they are expected to uphold an ethical obligation to ensure ChatGPT's utility and establish its credibility among users.

\paragraph{Hosting Platform's Responsibility} A subset of participants (5/12) identified hosting platforms
as responsible parties. They highlighted that these platforms are obligated to guarantee the ethical, legal, and secure deployment of these systems, emphasizing the platforms' financial interest in their business models. With P6 mentioned the hosting platforms \textit{``gonna be the people to decide how it operates and what it says"} and in contrast developers \textit{``don't have any control over it"}. This expectation is particularly pronounced when participants compare ChatGPT to widely used consumer products, with P3 stating, \textit{``[...] like if I choose to use eBay or Amazon, I expect them (the platform) to bear responsibility for the content they publish."}

\paragraph{ChatGPT Itself} Only P10 holds the view that the responsibility lies solely with ChatGPT. This perspective stems from their understanding that their engagement is \textit{``directly with ChatGPT"}, from which they receive information. Throughout this interaction, P10 does not take into account the potential involvement of any other entities. 

\paragraph{User Themselves.} Responsibility attribution to users themselves was less common
, with only 4 [P2, P3, P6, P7] participants acknowledging it. They cited a ``\textit{lack of sufficient knowledge}" about ChatGPT as a key factor limiting their effective engagement with the system. P3 reflected on this perspective, influenced by their awareness of friends using ChatGPT for advanced tasks, and concluded that their own limited understanding of generative AI technologies prevented them from fully exploiting ChatGPT's potential, stating, ``\textit{Nothing wrong with the machine. It's my lack of knowledge that stops me from getting the full experience.}"

\paragraph{Trust Changes after Receiving Deceptive Information} When it comes to participants perceived trustworthiness of the ChatGPT, especially when our researcher asked if their trust level had changed from the beginning participants start to use ChatGPT and after receiving the deceptive information. We first observed participants trust more along with their use of ChatGPT. To be more specific, initially, some participants [P2, P7] reported a low level of trust, influenced by negative perceptions or lack of understanding of ChatGPT. However, the trust increased later after recognizing its advantages [P2, P6, P7], and adjusting expectations accordingly [P10]. Conversely, others maintain a consistent level of scepticism or trust [P1, P3, P9]. For some, trust does not significantly change because they start with realistic expectations about the AI's accuracy and usefulness, particularly in their specific areas of interest or professional needs [P9, P10]. For example, P10 stated \textit{``I wasn't expecting it to be a hundred per cent right [...]"}. Four participants [P4, P8, P11, P12] experienced a decrease in trust when faced with inaccuracies or limitations in the responses. These experiences lead to a more cautious approach to using the technology, including verifying information independently [P11] and adjusting how they use the AI based on its limitations [P8, P12] (\textit{``[...], as trust goes down, I try to steer clear of using it for tasks where it's not doing great." [P12]}). 



\paragraph{Factors Affecting Trust.} Trust in ChatGPT, as reflected through participant feedback, is shaped by a complex interplay of factors that underscore the nuanced perceptions of its reliability and utility. 
\begin{itemize}
    \item Accuracy and Reliability [P2, P5, P9, P11]. Participants emphasized the critical importance of accurate and reliable information, with inaccuracies significantly undermining trust.
    \item Explanatory Transparency [P2, P7]. 2 participants highlight the value of clear explanations regarding ChatGPT's reasoning processes as a means to foster trust.
    \item Content Guidelines and Disclaimers [P5]: P5 pointed out that the implementation of content restrictions and disclaimers, especially on sensitive topics, informs users of ChatGPT's limitations, thus guiding trust (\textit{``from a trustworthiness perspective having disclaimer is important, you know what it can and cannot do [...], if you search for things about violence or suicide, it will decline to answer them"}). 
    \item Domain-Specific Trust Variations [P1, P2, P4-6, P8]: Trust varies significantly across domains, with participants expressing higher trust in specific areas (e.g., linguistic tasks) and caution in others (e.g., political content).
    \item Influence of External Perceptions [P10, P11, P12]: Participants expressed that narratives conveyed through media channels, as well as feedback from immediate social circles (including family and friends), have an impact on their level of trust towards ChatGPT.
    \item Clarity of Source Data and Integration with Tools [P7, P12]: The clarity regarding the data sources ChatGPT uses, and its integration with other tools, as mentioned by P7 and P12, are considered critical for building trust by providing insights into its operational mechanisms and utility. For instance, P12 said \textit{"if there are more and more tools build together with it, then I probably increase its usage frequency and trust."}.
\end{itemize}

\subsubsection{\textbf{Future Expectations of ChatGPT}}
\label{sec: future-expectations}
Participants shared suggestions and expectations for the future development of ChatGPT to get rid from deceptive behaviours from multiple perspectives.

\paragraph{General \& Technical Improvement Needs.} Participants underscored the necessity for advancements in ChatGPT to address and reduce incidences of deceptive behaviours and misinformation by enhancing transparency, and accuracy, and introducing robust validation processes. Furthermore, several participants expected ChatGPT to evolve beyond its current capabilities as a generative AI model. P8 acknowledged ChatGPT's potential to serve as a \textit{``real-time assistant"}, while P9 and 11 envisage it embodying more human-like qualities, facilitating its application across diverse life aspects, including education and healthcare sectors. P8 elaborated on these expectations by stating, \textit{``ChatGPT is capable of generating responses. However, what it lacks is the ability to act upon these responses. So I would like to see it not only generate responses but also execute actions based on them. That's where we're going to get to."}.


\paragraph{Verification \& Safeguard Strategies.} Participants in the discussion on verification and safeguard strategies for ChatGPT express a range of views emphasizing the shared responsibility between users and developers for verifying information accuracy. Foremost, it is argued that an enhancement of ChatGPT's capabilities to assess its certainty regarding the provided information and to authenticate the origins of its data prior to responding to users would be beneficial and should be prioritized. Many also advocated for the user's responsibility to perform due diligence [P8-11], while also recognizing the complexity of placing the entire burden of fact verification on developers. P11 specifically noted, \textit{``you [users]'re the one that agreed to their terms of service to use the platform so you always need to double check any information and should you take any information at face value."} This emphasizes the critical responsibility users bear in authenticating the information provided by the platform perceived by the participants. Furthermore, a notable preference for third-party verification emerges [P2, P3, P5, P6, P11, P12], suggesting it could offer unbiased, accurate checks and enhance user trust. However, participants also raised a subsequent concern regarding the credibility of the third-party verification tools themselves and the methods through which their trustworthiness can be substantiated (\textit{``So, because we don't fully trust ChatGPT, we're thinking of using a third-party tool for help. But then, do we need to double-check if that third-party tool is even reliable? It's like a never-ending loop."} [P1]).

\paragraph{Empowering Users \& Increasing their Self-esteem.}
This includes educating users, improving user self-protection, and continuous engagement in ChatGPT development and on in-using phase. Central to the dialogue is the imperative for enhanced educational initiatives and user guidance, 
to foster a deeper understanding of ChatGPT and the like: \textit{``I think everyone should have some kind of basic guidance or training on using ChatGPT before they use it [...] because it's important to know the type of technology you're using before you deploy it in the field." [P11]}. These considerations are of paramount importance, as highlighted in \S\ref{sec: deception-and-reaction}, for individuals lacking technological proficiency, including children and the elderly. Owing to their limited comprehension of ChatGPT's operational mechanisms, these groups are more vulnerable to adverse outcomes stemming from deceptive information.

The consensus among participants was that users need to be more aware of self-protection when using ChatGPT. Participants expressed the view that users should not place too much trust in ChatGPT, which still has limitations in understanding the nuances of individual situations and can be misleading. Therefore, they expressed concern and disapproval of relying on ChatGPT to make important life decisions: \textit{``I can't just go right on ChatGPT, can I sell my house for example or can I move to a different country? Those are decisions that need to be made by humans, not generative AI."} [P2]. In terms of the dependent on the use of ChatGPT, participants advocated a call for a cautious approach using ChatGPT, as they have concerns that it will destroy people's ability in critical thinking, and accordingly, lose their capabilities in specific tasks or skills. For instance, P12 mentioned that replying on ChatGPT will cause a more severely dependent on the ChatGPT and scarely, \textit{``Now when I write, I just let it go and I get lazy myself and don't deliberate on which word or phrase to use like I used to."} P8 also mentioned the importance of critical thinking and the ability to discern the credibility of information in the digital age, particularly in the context of social media and potential misinformation: \textit{``And if people start just trusting you at face value, because they've been told, oh, it's, it's vetted and it's regulated, then they kind of lose that ability to think critically for themselves. And it comes down to even on social media, if someone's posting an article with a misleading headline, like that skill of being able to verify and, you know, have your own critical thinking, it carries over into other aspects of life."}.

Participants pointed out that the current development and evolution of ChatGPT have not fully and seriously considered users' needs and experiences. Therefore, they advocated for a more inclusive, user-focused approach to the improvement of ChatGPT and other AI products. To be more specific, participants believe that it is crucial for companies to conduct user acceptance testing to ensure products meet the required standards for functionality and reliability (\textit{``if they don't get enough people saying, this is really good and it's producing accurate results, then they shouldn't release it" [P4]}). Once the product is in the market, collecting ongoing user feedback and learning from complaints and comments are vital for iterative improvements. P7 suggests that tools for feedback, such as \textit{``rating systems within the product"}, should have a tangible impact on development and enhancements. Furthermore, the discussion identified user studies as a crucial strategy to address significant gaps, including a limited understanding of user requirements and a neglect of ethical considerations. These studies are championed for their effectiveness in gaining a holistic comprehension of end-user needs, requirements, and feedback. P1 critically noted: \textit{``Without conducting user research, developers will just gonna think about everything from their or the company's angle, like how to make money.}"

\paragraph{Laws \& Regulations.} Except for P10, participants generally agreed on the necessity of enacting laws and regulations for the use of ChatGPT, especially considering its potential to generate deceptive information that could mislead users. Although P10 held a different view, their contention primarily revolved around the perceived urgency of implementing such regulations, which differed from others. P10 advocated for a wait-and-see approach, believing that, based on their and their associates' experiences, the use of ChatGPT was solely for entertainment purposes and thus did not necessitate elevation to a legal level. 
Moreover, concerns are raised about the rapid pace of technological advancement outstripping current legal frameworks, the government's slow response, and the potential for regulations to be either too intrusive or outdated [P4, P10-12]: \textit{``the technology is moving very fast and current measures and in place not really um up to date or quick enough to protect people"}. Simultaneously, participants [P3, P4, P10, P12] acknowledged the complexity inherent in achieving comprehensive global regulation of generative AI such as ChatGPT, given the backdrop of their escalating globalization. They pointed out that \textit{``not every country has the same level of expectation from the government, don't have the same safety expectations, etc" [P3]}, and  \textit{``I think we sometimes see that already with social media excuse me social media sites so we watch potentially legally in the US or another country might not be here and where does that quite stand and where is the date to hell." [P10]}. However, making some foundational rules that could be adapted and enforced locally could be the initial step. Additionally, as P11 mentioned \textit{``the government, they're not really experts in technology and AI" [P11]}, participants underscored the significance of fostering a collaborative regulatory approach that involves both governmental bodies and industry stakeholders, and suggesting that industry buy-in could accelerate the process of establishing relevant guidelines.


\section{Discussion and Summary of Results}
\label{sec:discussion}

In this section, we outline the major findings related to our research questions:
\begin{itemize}
    \item Survey participants primarily reported encountering \textit{over-simplified} responses whereas interview participants identified a new prevalent error: the \textit{partially correct} response. Furthermore, while the survey identified \textit{research} as the primary context for deception, interview data refined this to \textit{idea generation}. (RQ1)
    \item Survey participants' \textit{perceived responsibility} for deception in ChatGPT and their \textit{behavioural response} are influenced by a combination of personal and other factors. Conversely, \textit{worthiness} perceptions are solely impacted by the \textit{frequency of deception and use}. (RQ2)
    \item Interview participants expressed varied degrees of \textit{worthiness} in using ChatGPT for both personal and professional purposes. Opinions on \textit{responsibility} for ChatGPT's deployment diverged, with some attributing it to the hosting platform and others emphasizing the developers' critical role. Additionally, \textit{behavioural response} to deception differed, with some reducing their usage and others maintaining it, influenced by individual perceptions such as 
    their \textit{pre-existing domain knowledge of ChatGPT}, and \textit{trust}. (RQ2)
  
    \item Our chi-square analysis of the survey responses indicates that the \textit{frequency of deception} significantly impacts users' \textit{perceived worthiness of ChatGPT}, independent of personal factors. \textit{Responsibility} for deception is influenced solely by \textit{educational level} and \textit{frequency of deception}. In contrast, \textit{behavioural responses} are shaped by a combination of demographic variables (\textit{gender, age}) and other factors, including \textit{knowledge, verification tendency, and perceived worthiness}. (RQ3)
    
    \item Four interview respondents reported a decrease in trust towards ChatGPT post-deception illuminates the intricate nature of trust as a construct influenced by more than just deception encounters. The finding also identified additional factors impacting trust in ChatGPT, including \textit{explanatory transparency}, elaborated upon in \S\ref{sec: trust-and-responsibility}. (RQ4) 

    \item Interview participants expressed expectations for both general and technical enhancements in ChatGPT to mitigate deception, particularly emphasizing the potential of improved \textit{explanations}. A significant portion advocated for user-centric approaches, suggesting that future design and development should more thoroughly incorporate user needs and feedback. Despite acknowledging potential challenges, participants also engaged in a meaningful discussion on the implementation of laws and regulations as viable strategies. (RQ5)
\end{itemize}

We proceed to discuss the results of our study in more detail.

\subsection{Integrated Impact of Mental Models, Experiences, and Personal Factors}

The results from the second part of our study that focused on interviewing participants (Study II), highlighted how participants' mental models, experiences (as reported in \S\ref{sec:mental-model}), and personal factors influence their perception of ChatGPT's deception, and how it affects their trust and changes in their behaviour. 

For instance, individuals who do not impose high expectations and standards on ChatGPT for fulfilling their specific requirements or for application in serious contexts tend to maintain a generally relaxed disposition throughout the interview process. As presented in \S\ref{sec: deception-and-reaction}, in areas deemed trivial by the users, there is often no effort to verify the accuracy of information provided by ChatGPT. 

Furthermore, there is a tendency among these individuals to blindly trust the information furnished by ChatGPT, often motivated by personal curiosity and the influence of social conformity. In contrast, individuals who approach ChatGPT with the intention of obtaining meaningful and valuable information for task execution exhibit a markedly different interaction pattern. These participants seem to engage in a deeper level of critical thinking concerning the credibility of ChatGPT. The participants (i) actively contemplate the underlying reasons for any discrepancies or errors that they encounter, (ii) deliberate on who might be accountable for these issues, and (iii) consider potential measures for resolution.

The rigorous approach of utilizing ChatGPT in this way reflects what can the behaviours observed in privacy-related research. This suggests that there might be an underlying stratification among users based on their concerns and understanding of privacy. For example, prior studies~\citep{hsu2022privacy,chen2023research,soumelidou2021towards}  have identified that while some individuals are indifferent to the sharing or access of their information by others online, those who prioritize privacy as a paramount concern demonstrate a heightened awareness and demand for privacy safeguards. 


Our study found that participants reported varying levels of knowledge about ChatGPT, ranging from moderate to expert, as detailed in \S\ref{sec:mental-model} and supported by data in Table~\ref{tab: demo-interview}. However, when asked more detailed questions, we discovered that many participants could not provide thorough or accurate explanations about how ChatGPT works or why ChatGPT might give deceptive answers. This indicates a gap between what participants think they know about ChatGPT and their actual understanding. This knowledge gap challenges the reliability of some of our study's findings, including those from the Chi-square test. 

Moreover, our study suggests that people's understanding of ChatGPT's deception is linked to their beliefs about responsibility and knowledge. Specifically, some participants believe that users themselves are responsible for the deceptive responses given by ChatGPT. They argue that the quality of ChatGPT's replies depends on the quality of the information or questions users provide. If the inputs are poor or misleading, ChatGPT's responses will be sub-optimal, suggesting that better user input could improve ChatGPT's performance.



Overall, our study elucidates that personal factors influencing user interaction with ChatGPT extend beyond mere demographics to encompass professional affiliations, job-related experiences, and personal traits that participants disclosed during interviews. For instance, some users described themselves as inherently trusting individuals, while others, particularly those with professional ties to the AI industry or who have witnessed numerous AI failures, expressed a heightened sense of caution towards using ChatGPT. These observations underscore the complexity of user engagement with AI tools and highlight the necessity for a user-centric approach in this domain. We argue that adopting such an approach is not only critical but also in urgent demand within the field, necessitating significant attention from corporations. 


\subsection{User-centric Approaches against Deception}
\label{sec:user-drive-approach-discussion}

This study underscores the critical need for user-centered research and methodologies in developing language model technologies, particularly to address and mitigate deception in ChatGPT and other LLMs. Such an approach is not only essential for overcoming current deception challenges but is also vital for the ongoing improvement and development of these technologies.

The rapid advancement and deployment of LLMs, driven largely by commercial incentives, often overlook crucial ethical considerations, including the risk of deception. This oversight poses significant risks, especially to vulnerable groups, as highlighted in our findings. The pursuit of profit at the expense of ethical integrity underscores the need for a comprehensive framework to address these concerns.

User-driven research offers a twofold benefit. First, it ensures that enhancements are informed by actual user interactions and experiences, leveraging real-world insights over theoretical assumptions. This is increasingly relevant as LLMs evolve to adapt to user preferences through reinforcement-learning-through-human-feedback (RLHF) strategies. Direct user feedback becomes a cornerstone for refining these models, improving both their accuracy and ethical alignment.

Secondly, prioritizing users fosters an ethical development culture, where potential harms are identified and addressed proactively, integrating ethical considerations from the outset. This approach not only helps to prevent harm but also rebuilds user trust, showcasing a commitment to user welfare beyond mere profit.

In summary, the transition towards user-centric methodologies in the development of ChatGPT and similar LLMs is imperative for ethical, responsible AI development. This shift emphasizes the balance between innovation and ethical responsibility, ensuring the creation of technologies that are not only advanced but also safe and aligned with societal values. Our study highlights the urgency of this transition, advocating for a development paradigm that equally values user insights and ethical standards.


\subsection{Navigating the Trade-off: User Reliance on ChatGPT and the Need for Verification}
\label{sec: trade-off-discussion}

We observed that there seems to be a notable discrepancy between the anticipated convenience offered by the utilization of ChatGPT and the actual experiences of users. This discrepancy emerges from the findings detailed in \S\ref{sec: deception-and-reaction}, \S\ref{sec: trust-and-responsibility}, and \S\ref{sec: future-expectations}. These results highlight participants' concerns regarding the accuracy of information provided by ChatGPT. Participants found themselves compelled to employ additional measures and exert significant effort to verify the correctness of the information provided by ChatGPT, which seems to contradict the initial rationale behind employing ChatGPT, i.e. the rationale that ChatGPT will make their job easier. As participants have indicated, this verification process often results in a recursive loop where assistance sought from ChatGPT ends up being cross-checked with Google or other databases, raising the question: \textit{why not directly utilize Google or a reliable search engine in the first place?}

On the other hand, it is important to acknowledge that users do indeed derive considerable benefit from ChatGPT's capability to aggregate information. This is especially the case in the preliminary stages of performing specific tasks such as identifying research directions or sourcing informational pathways. This contribution is perceived as remarkable by users. The trade-off between the convenience offered by ChatGPT and the need for additional verification measures constitutes a critical area for discussion. 

In response to this contradiction, potential solutions involving the implementation of additional safeguards and fact-checking tools were proposed and discussed. These tools aim to alleviate the time and extra burden associated with verifying ChatGPT-provided information. While users expressed appreciation for these measures, their responses also reflected a cautious approach towards relying on third-party tools, driven by a desire for personal involvement in the verification process to retain a certain degree of control. This cautiousness from participants suggests that they are not inclined towards extreme measures — i.e. neither relying entirely on third-party tools without performing any verification themselves nor consistently conducting their own fact-checking each time they receive a response. Instead, the prevailing trend for our study participants appears to favour a balanced approach that incorporates the use of verification tools while maintaining user control over the process.

\subsection{Limitations}
Our research is based on self-reported data, which may possibly overlook and not fully capture the complexity of users' experiences and perceptions. Additionally, the rapid evolution of AI technologies like ChatGPT means that user perceptions and the platform's capabilities could change, potentially dating our findings. Future research should consider longitudinal studies to better track changing user perceptions and AI advancements. Expanding investigations to assess regulatory and educational measures' effectiveness against deception and developing user-friendly AI verification tools are crucial next steps. 


\section{Related Work}
\label{sec:rel_work}


 In this section we discuss the related work at the intersection of Deceptive AI \& Society \citep{sarkadi2023deceptive}. Deceptive AI and society research ranges from the more recent studies that try to figure out the relation between LLMs and deception, to the original idea of building a socio-cognitive theory of trust and deception proposed by \cite{castelfranchi2001trust}.


The doubt of whether a given technology can be wholly beneficial without any accompanying drawbacks still persists. While explicit failures are readily observable by human eyes, implicit errors are harder to identify and/or fact-check. On one hand, it may be tolerable for chatbots to generate nonsensical responses that merely frustrate users. On the other hand, the possibility of ChatGPT producing \emph{misleading} and \emph{deceptive} information is a matter of serious concern~\citep{susanlie,Tiffany}, especially if adopted on a large scale to offer services to users or in safety-critical systems. 

Deceptive information through LLM `hallucination' can have adverse impacts on users who are not equipped to distinguish `fact' from `fiction'~\citep{rohrbach2018object,xiao2021hallucination}. The dangers that ensue from the use of banal deceptive AI range from LLM being used as tools by other humans to manipulate individuals to causing real harm, and in the extreme, may even result in broader societal ramifications, such as a lack of shared trust among community members and governmental institutions.

A significant contribution in the area of chatbot technologies is the empirical research conducted by ~\cite{mcguire2023reputational}, who examined user reactions to deceptive behaviors in chatbots. Their findings suggest that users can often fail to recognize deceptive cues, leading to misplaced trust in AI systems. Similarly, ~\cite{ehsan2021expanding} focus on the impact of transparency mechanisms in mitigating the effects of trust, indicating that clear communication about an AI's capabilities and limitations can enhance user discernment.


\citeauthor{pacchiardi2023catch} specifically addresses the challenges posed by LLMs, including their ability to generate plausible yet factually incorrect or misleading information. This study emphasizes the need for improved detection mechanisms and develops the detector works by asking a predefined set of unrelated follow-up questions.

Going back to balancing our doubts about the threats and benefits of deceptive AI technology in society, we must also emphasise ongoing research that aims to delve deeper into how such technologies can be beneficial and how human-AI interactions work.

In the sub-area of AI called argument mining, a line of work of has been to detect deceptive arguments in politcal debates and contexts using BERT-style systems \citep{delobelle2020sifting,goffredo2023disputool}. A similar line of work has focused on identyfying fallacies and hate speech \citep{goffredo2023argument}.


The study of human perceptions of deceptive AI behaviour has been covered in various domains. The types of behaviours have mainly been considered in linguistic, motor, and social contexts.

In human-robot and human-AI interaction (HRI and HAI), \cite{dragan2014analysis} explored how humans perceived the presence of deceptive intentions based on pre-calibrated motions of robotic arms. In the same research area, \cite{chakraborti2019can} studied how acceptable AI generated lies were in human-AI teaming search \& rescue scenario. Furthemore, \cite{sarkadi2023should} explored how AI agents are perceived compared to humans when it comes to job roles that involve deception in various agent-agent interactions, including human-AI teaming. Human-AI interaction scenarios also involve the phenomenon where different AI agent strategies can increase human willingness to deceive \citep{mell2018welcome}. In particular, such effects can be observed in Human-AI negotiation settings \citep{mell2020effects,jahan2024unraveling}.  

Two crucial abilities of deceptive AI technologies missing from LLMs are that of reasoning and planning \citep{sarkadi2021phdthesis,valmeekam2023planning}. In the area of AI planning research, an important line of work has looked at extended goal recognition \citep{masters2021extended} and deceptive path planning \citep{masters2017deceptive,price2023domain}. Regarding reasoning, \cite{sarkadi2019dectom} have explored how AI agents can use abductive and practical reasoning with ToM to cause desired false beliefs in other agents. In the area of multi-agent reinforcement learning (MARL), deceptive AI has been studied in different setups. \cite{piazza2023theory} looked at how localised models of Theory-of-Mind can be used to distinguish between cooperative and deceptive AI agent communication.

Another area where Deceptive AI has been studies is in the one of Artificial Societies and Simulation. Specifically related to Deceptive AI \& Society, the work of \citeauthor{sarkadi2021evolution} has shed light how deception evolves in human-AI agent societies \citep{sarkadi2021evolution}; how these societies can self-organise in the face of deception to re-establish cooperative communication through social learning and System 2 type of critical thinking and investigation mechanisms \citep{sarkadi2024self}; how the presence of deception triggers an arms-race in Theory-of-Mind between deceivers and investigators \citep{sarkadi2023arms} - a similar result is observed in the MARL approach where ToM is used as a model for Inverse Reinforcement Learning proposed by \cite{alon2023dis}; and how competition between agents creates evolutionary pressures to make deception a stable strategy \citep{sarkadi2024triangles}.

Deceptive AI research is only starting to gain traction as a subarea of AI. Several workshops on the topic have been organised in the past years, including the Deceptive and Counter-Deceptive Machines AAAI Fall Symposium\footnote{\url{https://aaai.org/proceeding/03-fall-2015/}}, the Machine Deception Workshop at NeurIPS\footnote{\url{https://www.machinedeception.com/}}, the 1st and 2nd International Workshops on Deceptive AI co-located with ECAI 2020 and IJCAI 2021\footnote{\url{https://sites.google.com/view/deceptai2021}, see proc. in \citep{sarkadi2021deceptivebook}}, and more recently the Rebellion and Disobedience of Artificial Agents Workshop Series co-located with AAMAS \footnote{\url{https://sites.google.com/view/rad-ai/home}}. 

Overall, there's a common thread in all Deceptive AI research, namely that of aiming to better capture what deception is in relation to AI technology, and how it can be used for the good of society rather than increasing risk. As \cite{coeckelbergh2018describe} notes, understanding deceptive AI is not just about the technology and engineering of computational systems, but about an overarching narrative about the politics of technology, the power relations and structures that drive technology, and, last, but not least, how these play into human cultures and psychological biases. In other words, Deceptive AI needs to be understood as part of an ecosystem \citep{zhan2023deceptive}.

\section{Future Directions for Artificial Minds}
\label{sec:future_dir}

\begin{quote}
    \textit{As Artificial Intelligence (AI) technology advances, we increasingly delegate mental tasks to machines.  However, today's  AI systems usually do these tasks with an unusual imbalance of insight and understanding...}\citep{lewis2024reflective}
\end{quote}

We are in the AI age of `\textit{incomplete minds}' \citep{lewis2024reflective}. This is also the case regarding the task of communicating or engaging in dialogue with others. Machines designed and pre-programmed by humans to pass the Turing test, the so-called chatterbots, follow a pre-defined script given to them by their designers \citep{mauldin1994chatterbots}.  In the case of Large-Language-Models (LLMs), these follow a statistical distribution of word embeddings to generate textual output \citep{bender2021dangers}. Both types of machines have one thing in common, namely that they do not posses the ability to deliberate in order to decide what to say, how to say it, or when to say it in different circumstances. Such machines are either (i) vessels of deception, performed by humans through script design or language bias propagation, or (ii) `bullshit' machines that are completely ignorant of the truth-value of their utterances \citep{hicks2024chatbullshit}. Yet, if placed in particular contexts, they can trigger humans into extreme cases of anthropomorphisation \citep{natale2021deceitful,sarkadi2023deceptive}. 

A subtle thread throughout this paper is that of \textit{worthiness} of interaction. In the particular context of communicative AI agents, this means the worthiness of talking to such an AI agent. Hence, a pertinent question to ask is the one Charles Hamblin hinted at, namely \textit{How do we build a machine worth talking to?}. To do this, Hamblin was one of, if not the first in the area of AI, to propose the mathematical modelling of dialogue \citep{hamblin1971mathematical}, and finally the design of an AI agent architecture that enables the modelling in a similar fashion of the mind of the AI agent's interlocutor \citep{staines2018linguistics}.

Instead of applying script-based policies \citep{schank1975scripts} or `parroting' algorithms \citep{bender2021dangers} for tricking humans into believing they are engaging in meaningful interactions, there might be a future where machines are given cognitive abilities that humans use to govern their communicative behaviour, think about future consequences, and learn from simulating future consequences of their communicative acts before they actually perform those acts. One such ability is that of reflection \citep{lewis2024reflective}. 

Reflective machines might be created to actually be able to use their abstract models of the world and others to give semantics to their utterances or even to their non-linguistic behaviour, similarly to the dialogical agents proposed by \cite{mcburney2002rational}, or the deceptive AI agents proposed by \cite{sarkadi2021phdthesis}, which have internal `consequence engines' that simulate the outcomes of communicative interactions with respect to the false beliefs formed in the minds of their interlocutor agents. These sorts of agents not only have the ability to model other agents behaviourally, as explored in the special issue edited by \cite{albrecht2018autonomous}, but have the ability to use an Artificial Theory of Mind to reason about consequences of their actions on the minds of others. For instance, the agents in \cite{sarkadi2021phdthesis} use a combination of simulated theory of mind (ST) and theory-theory of mind (TT) to reason about how they can cause changes in the beliefs of others and reason about the consequences of these belief changes, albeit on a high-level. Similarly, Winfield's robots use it to predict the actions of other agents and anticipate the likely consequences of those actions both for themselves and the other agents~\citep{winfield2018experiments}.

What is the actual \textbf{research challenge} in AI here? When AI agents talk about things in a human-interpretable manner they need to make sense, not talk nonsense, blabber, or bullshit. First, work on speech-acts and agent communication languages has to be further developed to enable agents to extract and refer to linguistic semantics from their abstract models of the world \citep{cohen1995communicative}. Second,  methods based on natural-language-processing and argumentation \citep{cabrio2012natural, lawrence2020argument} need to be developed for agents to be able to perform Abstract Conceptualisation from linguistic or other types of data. Finally, there is need is to integrate dialogue-based argumentation frameworks \citep{McBurney-2009-DGforAA} for reflective agents to be able to form sound and consistent arguments, and even tell meaningful stories when interacting with others, without resorting to pre-defined scripts \citep{schank1975scripts} or learnt statistical patterns.

One way forward to address this challenge is to first adopt a neurosymbolic approach to AI system design \citep{garcez2023neurosymbolic}. Yet, neurosymbolic models aren't enough on their own. An architecture that not only integrates mutliple neurosymbolic models, but lets them evolve in an interpretable manner through reflection is required, \textit{à la} \textit{Reflective AI} \citep{lewis2024reflective}. Communicative agents need to be able to reason about the consequences of their actions, be aware of their world, e.g. environment or society, be self-aware of their knowledge and capabilities, about others' knowledge and capabilities, i.e. be able to use Theory of Mind, and reflect on both what they and their interlocutors utter. Most importantly, as pointed out by \cite{isaac2017}, when it comes to distinguishing between malicious and pro-social deception, AI agents must be able to reason about the ethical values of their interlocutors, and at least try to align themselves to those ethical values. AI agent architecture is crucial for enabling communicative agents to do these things. At the same time we need to ask ourselves about what processes are missing in current AI systems and what processes are needed to be included in future AI architectures.

\section{Conclusion}
\label{sec:conclusion}

In this study we aimed to partially elucidate the complex landscape of human perceptions toward deception when using LLM-based AI technologies, such as ChatGPT. Our two studies bring insights into the types of deceptive information encountered and the contexts of their occurrence. We highlighted the critical impact of deceptive behaviors on user trust and the varied responses individuals exhibit towards perceived deceptive information. Notably, our findings re-emphasise the need pointed out by \cite{castelfranchi2002role} more than 20 years ago, namely that we need a multi-dimensional socio-cognitive approach to address both trust and deception in human-AI societies. According to our study, this approach should nowadays consider user education, technical improvements, and robust regulatory frameworks. Furthermore, the study calls for continued exploration into user-centric methodologies and the development of ethical AI systems that prioritize user welfare. Our work aims to set a foundation for \textit{Deceptive AI Ecosystems} by navigating the trade-offs between AI convenience and the need for verification, in order to inform the creation of more transparent, accountable, and trustworthy AI technologies.

Yet, as part of our Deceptive AI Ecosystems approach, these technologies remain just one element of a bigger problem in designing trustworthy AI systems. We should probably to change how we talk about AI and deception. An important factor in Deceptive AI Ecosystems, that enhances the deceptiveness of AI technologies, is the creation of context around it and the reinforcement of biases by using AI as a speech act with the ulterior goal of monetisation of individuals and society \citep{lewis2021ai}. The trend of democratic backsliding is enhanced not only by the use of technologies, but also by the way in which we communicate and `normalise' our perceptions and beliefs about these technologies, which in today's techno ecosystem is done through social influence by actors/agents, such as BigTech, who have both the power and incentive to do so \citep{mertzani2022social}. It is at this level where intentional deception happens, rather than at the technological `stochastic parroting' level. \cite{coeckelbergh2018describe} is right in the sense that deceptive AI is not just baout the technology, but as \cite{zhan2023deceptive} point out, it's also about the ecosystem, and we must always keep this in mind when evaluating Deceptive AI technologies. LLMs do not have an incentive to deceive, but improperly regulated human-led organisations do, especially when the global cultural market paradigm promotes market competition at the expense of human values. 

\begin{sloppypar}
  As shown by \cite{sarkadi2024triangles}, competitive contexts provide the ideal ecosystem for deceptive behaviour to become evolutionarily stable. Looking at this phenomenon from a Cybernetic \citep{wiener2019cybernetics} and Techno-Political \citep{pitt2021self} perspective of human-AI ecosystems, we can notice that the large-scale deployment of AI technologies and the LLM arms-race\footnote{\seqsplit{https://www.economist.com/leaders/2024/05/16/big-techs-capex-splurge-may-be-irrationally-exuberant}accessed on 30th May 2024} between Big Tech will lead to an optimal ecosystem for deceptive behaviour. Hence, our best hope as a truth-seeking AI community in this competitive technological context is to take the role of investigators in the mentalisation arms-race against agents of malicious deception \citep{sarkadi2023arms} whilst looking for ways to promote pro-social AI deception \citep{castelfranchi2000artificial}.  
\end{sloppypar}

\section*{Acknowledgments}

XZ and YX formulated the research questions designed the study, ran the experiments and performed the statistical analysis. SS formulated the research questions and supervised the study. All authors contributed to study design and the writing of the paper.



\bibliography{references}
\bibliographystyle{apalike}

\end{document}